\documentclass[aps,prl,reprint,amsmath,amssymb,nofootinbib,superscriptaddress]{revtex4-2}

\usepackage[T1]{fontenc}
\usepackage{lmodern}
\usepackage{microtype}
\usepackage{mathtools}
\usepackage{bm}
\usepackage[dvipsnames]{xcolor}
\usepackage{hyperref}
\hypersetup{colorlinks=true,linkcolor=MidnightBlue,citecolor=MidnightBlue,
            urlcolor=RoyalBlue}

\newcommand{\cR}{\mathcal R}
\newcommand{\Om}{\Omega_{\rm GW}}
\newcommand{\kbh}{k_{\rm BH}}

\begin{document}

\title{Causal Gravitational-Wave Production Does Not Generate Infrared
Adiabatic Curvature}

\author{Ruth Durrer}
\email{ruth.durrer@unige.ch}
\affiliation{Department of Theoretical Physics, University of Geneva,
24 quai Ernest-Ansermet, 1211 Geneva 4, Switzerland}

\author{Alessandro Melchiorri}
\email{alessandro.melchiorri@uniroma1.it}

\affiliation{Physics Department, Sapienza University of Rome,
Piazzale Aldo Moro 5, 00185 Rome, Italy}

\begin{abstract}
Short tensor modes can generate a white noise contribution to the local scalar
$K=8\pi G\rho/3-\theta^2/9$, but this does not induce an infrared-enhanced
adiabatic curvature spectrum. For a conserved causal source, analyticity of the
longitudinal velocity correlator requires $P_V=\mathcal O(k^2)$, hence
$P_{\mathcal R}=\mathcal O(k^0)$ and
$\Delta_{\mathcal R}^2=\mathcal O(k^3)$, excluding the
$\Delta_{\mathcal R}^2\propto k^{-1}$ relic. We also construct a conserved
conversion of radiation into gravitational waves: the white field resides in a
compensated entropy mode, and the fluid response cancels the slowly decaying
$a^{-2}$ relic. Adiabatic bounds on such sources are therefore source- and
matching-dependent.
\end{abstract}

\maketitle

Nonlinear mode coupling transfers power from short to long wavelengths, so
causal microphysics in the early Universe can leave a white power spectrum,
$P(k\to0)=\mathrm{const}$, on cosmological scales. What matters is which
variable is white: since $P_X(k\to0)=\mathrm{const}$ gives
$\Delta_X^2\propto k^3$, a white spectrum for a density-like variable does 
imply a $k^{-4}$ spectrum for its inverse-Laplacian potential. However, the observable response is fixed
by the total constrained system and by the history generating it, not by any
isolated contribution.

A recent proposal \cite{PaperI,PaperII,BS2026} isolates the contribution of
short tensor modes to
\begin{equation}
K \equiv \frac{8\pi G}{3}\rho - \frac{\theta^2}{9}.
\label{eq:Kdef}
\end{equation}
Convolving the short modes gives a white spectrum for the isolated transverse
contribution, $P_{K_\perp}(k\to0)=\mathrm{const}$. If the curvature $\cR$ is related to $K_\perp $ through a linear
Poisson-type equation, this is then mapped to an infrared curvature mode,
\begin{equation}
\Delta^2_{\cR}(k) = \Delta^2_{\cR,\rm ad}(k) + \frac{\kbh}{k}\,.
\label{eq:template}
\end{equation}
which is Eq.~(1) of Ref.~\cite{BS2026}, with $\kbh$ defined there as the
$k\to0$ limit of $k\,\Delta^2_{\delta\cR}$ in its Eq.~(10).

The two terms of Eq.~\eqref{eq:template} are logically distinct: white noise in
$K$ can arise without inducing an inverse-Laplacian adiabatic mode in the curvature. We establish this
first from causal analyticity, and then by constructing an explicit compensated
tensor-production history.

\textit{Causality and analyticity.---} Consider scalar perturbations of an ideal fluid
with constant equation of state $w=c_s^2$ and negligible anisotropic stress. On
superhorizon scales the regular growing solution has constant gravitational
potential $\Psi$, and the velocity constraint reads \cite{DurrerBook}
\begin{equation}
 \frac{3(1+w)}{2}\mathcal H^2 V
 = k\left(\mathcal H\Psi+\Psi'\right),
\label{eq:velocityconstraint}
\end{equation}
where $v_i=i\hat k_iV$ and $\mathcal H=a'/a$. Since
$\mathcal H=2/[(1+3w)\eta]$, the growing mode obeys $V=C_w(k\eta)\Psi$ with
$C_w=(1+3w)/3(1+w)$, and the comoving curvature $\cR$ differs from $\Psi$ only
by a finite, $k$-independent factor in this limit.

Suppose now that the scalar velocity field is generated by a causal process at
$\eta_*$, with no pre-existing superhorizon correlations. Its real-space
two-point correlation function then has finite range, so its Fourier transform
is analytic at $\bm k=0$. Statistical isotropy gives for the longitudinal part
\begin{equation}
 \langle v_i(\bm k)v_j^*(\bm k')\rangle
 = (2\pi)^3\delta_D(\bm k-\bm k')\,
 \hat k_i\hat k_j P_V(k).
\label{eq:velocitypower}
\end{equation}
Because the projector $\hat k_i\hat k_j=k_i k_j/k^2$ is nonanalytic at the
origin, analyticity of the full correlator requires
$P_V(k)=v_2 k^2+\mathcal O(k^4)$, unless further cancellations make the
spectrum steeper still. With the growing-mode relation this gives
\begin{equation}
 P_{\cR}(k)=\mathcal O(k^0),
 \qquad \Delta_{\cR}^2(k)=\mathcal O(k^3).
\label{eq:causalscaling}
\end{equation}
Equation~\eqref{eq:template}, by contrast, requires $P_{\cR}\propto k^{-4}$ and
hence $P_V\propto k^{-2}$, which is incompatible with a causally generated
regular velocity field: a nonzero coefficient of that branch must be supplied
as independent superhorizon initial data. The argument concerns the free
regular scalar mode after the source has switched off; active anisotropic
stress and relative-entropy modes require their own transfer functions.

One qualification points directly to the construction below. A gravitationally
induced velocity field is fixed by elliptic constraints, so finite source range
does not by itself guarantee an analytic correlator: what enforces
this scaling is local conservation, through the causal
integral constraints \cite{Traschen1985}, and seeds with unsuppressed scalar
anisotropic stress can evade the naive counting
\cite{DurrerSakellariadou1997}. Our two arguments are therefore complementary
rather than logically independent.

The same distinction appears at the level of the tensor source. Only the total
$K$ enters the constrained scalar geometry: its transverse, scalar, and mixed
pieces cannot be assigned independent initial data. Grouping every piece other
than the isolated transverse one into $K_r=K-K_\perp$,
\begin{equation}
P_K = P_{K_\perp} + P_{K_r} + 2P_{K_\perp K_r}.
\label{eq:crossterm}
\end{equation}
The local process that creates the waves also perturbs the parent fluid, so the
cross spectrum does not generically vanish; causal-seed calculations provide
familiar examples in which the fluid response compensates a white source on
superhorizon scales \cite{VS1990,HuSpergelWhite,DurrerSakellariadou1997}. A
positive spectrum for an isolated contribution therefore does not imply one of
the same size for the total. We now give an example in which local
conservation enforces the anti-correlation.

\textit{Compensated perturbations.---} We adopt a two-scale description. Short
tensor waves are coarse grained into an effective radiation component indexed $g$,
while their scalar envelope and the parent radiation $r$ vary on much longer
wavelengths---the usual regime in which the averaged wave stress tensor obeys
$\bar p_g = \bar\rho_g/3$ \cite{Isaacson1968}, realized by a locally isotropic
ensemble of transverse-traceless wave packets with variance chosen to reproduce
$\rho_g(\bm x)$. The split
$\nabla_\mu T^{\mu\nu}_g = Q^\nu = -\nabla_\mu T^{\mu\nu}_r$ conserves the
total stress tensor exactly.

Let $s(\bm x)$ describe fluctuations in the number or strength of localized
production events, with effective causal correlation scale
$\ell_c \lesssim H_*^{-1}$ and $k_c \equiv \ell_c^{-1}$. For randomly placed
events, causality gives
$P_s(k) = P_0[1 + \mathcal O(k^2/k_c^2)]$ for $k \ll k_c$. To leading order in gradients, a finite-duration energy exchange is
specified by 
\begin{equation}
\left[a^4\delta\rho_g\right]' = A\,W'(\eta)\,s(\bm x),
\qquad
\left[a^4\delta\rho_r\right]' = -A\,W'(\eta)\,s(\bm x),
\label{eq:exchange}
\end{equation}
with $W(\eta)$ a smooth window rising from zero to unity during
$\eta_* < \eta < \eta_*+\Delta\eta$.
No energy is created: what is deposited in the
gravitational-wave component is removed locally from the parent radiation. Once
the source switches off, $\delta\rho_g = A a^{-4}s = -\delta\rho_r$.

We take the emission to be isotropic in the parent-fluid rest frame at the
coarse-graining scale. Since both components have $w=1/3$, compensation and
isotropy imply the $k$-scaling for $k<k_c$
\begin{equation}
\begin{gathered}
\left\{\delta\rho,\,\delta p,\,\pi\right\}_{\rm tot}
= \mathcal O\!\left[(k/k_c)^2\right]\delta\rho_g , \\
q_{i,\rm tot} = ik_i q_1 + \mathcal O(k^3)
= \mathcal O\!\left[(k/k_c)\right]\delta\rho_g ,
\end{gathered}
\label{eq:compensated}
\end{equation}
 The
momentum enters at one power of $k$, not two: $ik_iq_1$ with
$q_1=\mathcal O(k^0)$ is the general isotropic form of a longitudinal vector,
so statistical isotropy does not remove it. For the explicit counterexample we
adopt the stricter common-velocity realization and set $q_1=0$ during
matching---an assumption of that realization, not a consequence of isotropy,
whose cost we return to below. Starting from an unperturbed scalar geometry,
the Hamiltonian and momentum constraints then admit
\begin{equation}
\begin{gathered}
k_{\rm BH}=0, \qquad \Phi = \Psi = \cR = K = 0 \\
\text{(strict zero-transport limit)},
\end{gathered}
\label{eq:vanishing}
\end{equation}
which persists after production in the perfect-radiation limit, the two
components obeying identical scalar evolution equations.

The qualification is essential, and what survives beyond it is the point of the
construction. The local relation $\delta\rho_r=-\rho_{\rm GW}$ used below
defines the compensated history; it is what conservation permits, not what it
enforces for an arbitrary emitter. And the waves propagate away from the
deficit they left behind: during the finite production and matching interval
the transport distance is $\lesssim\ell_c$, so corrections begin at
$\mathcal O[(k/k_c)^2]$, while at later times the relevant transport scale grows
and the resulting evolution belongs to the source-dependent transfer problem.
Compensation thus suppresses the monopole rather than completely removing the
perturbation: with
$\delta\rho_{\rm tot}=\mathcal O[(k/k_c)^2]\delta\rho_g$, the Poisson equation
$k^2\Phi\propto a^2\delta\rho_{\rm tot}$ leaves a potential that is finite and
white at finite $k$, $P_\Phi=\mathcal O(k^0)$, smaller by $(k/k_c)^2$ in
amplitude than for an uncompensated source. Equation~\eqref{eq:vanishing}
therefore excludes the infrared-enhanced branch, but not a finite causal scalar
response. That regular white potential is the branch of
Eq.~\eqref{eq:causalscaling}, reached here from an explicit conserved history
rather than from analyticity: the two arguments agree on the same power, and do not lead to the $P_{\cR}\propto k^{-4}$ branch required by
Eq.~\eqref{eq:template}.

The perturbation has not disappeared. Writing $\delta_i=\delta\rho_i/\bar\rho_i$,
it resides in the relative entropy
\begin{equation}
S_{gr} = 3(\zeta_g-\zeta_r) = \frac{3}{4}(\delta_g-\delta_r) \neq 0,
\label{eq:entropy}
\end{equation}
whose spectrum inherits this white tail: the post-source state is a compensated
isocurvature mode, not the adiabatic mode of Eq.~\eqref{eq:template}.

\textit{Covariant matching.---}The effective-fluid construction absorbs the
short-scale shear into $\rho_g$. To check that this has not removed the effect
by assumption, we recast the same matching in the fine-grained covariant
variables of Ref.~\cite{BS2026}, retaining the shear explicitly. Overdots now
denote cosmic-time derivatives and $H=\dot a/a$.

In the fine-grained description the wave energy enters through the shear rather
than through $\rho$, and neglecting vorticity the Raychaudhuri equation reads
\begin{equation}
\delta\dot\theta + \frac{2}{3}\bar\theta\,\delta\theta
= -2\,\sigma_\perp^2 - 4\pi G\,\delta(\rho+3p)_r .
\label{eq:raych}
\end{equation}
The energy carried by the waves is identified as
$\rho_{\rm GW}=\sigma_\perp^2/(4\pi G)$---the effective tensor-energy
identification adopted in Ref.~\cite{BS2026}, which presupposes an average over
many wave oscillations \cite{Isaacson1968}---and by construction it is removed
locally from the parent radiation, $\delta\rho_r=-\rho_{\rm GW}$. For $w=1/3$
this gives $\delta(\rho+3p)_r = 2\delta\rho_r = -\sigma_\perp^2/(2\pi G)$, so
the two sources in Eq.~\eqref{eq:raych} cancel identically,
$-2\sigma_\perp^2+2\sigma_\perp^2=0$, and $\delta\theta=0$. The extra
deceleration produced by the shear is offset by the energy deficit left
in the fluid. As in Eq.~\eqref{eq:vanishing}, this result is not exact but
holds up to local causal motions accompanying production, which are
gradient suppressed by $(k/k_c)^2$; the homogeneous solution
$\delta\theta\propto a^{-2}$ is excluded by the unperturbed pre-production
geometry. After production both terms redshift as $a^{-4}$, so the cancellation
persists; if $\delta\rho_r=-\rho_{\rm GW}$ is imposed throughout the
conversion, it holds during the finite source interval as well.

With $\delta\theta=0$, the total $K$ follows from its definition,
\begin{equation}
K_{\rm tot} = \frac{8\pi G}{3}\delta\rho_r = -\frac{2}{3}\,\sigma_\perp^2
\;\propto\; a^{-4} .
\label{eq:Ktot}
\end{equation}
Here $\delta\rho_r$ appears rather than $\delta\rho_{\rm tot}$ because in the
fine-grained description $\rho$ in Eq.~\eqref{eq:Kdef} is the parent fluid
alone, the wave energy being carried by $\sigma_\perp^2$; coarse grained, that
same energy sits in $\rho_g$, giving $\delta\rho_{\rm tot}=0$ and $K=0$. The
two bookkeepings differ in where the wave energy is entered, not in the
physics. Like Eq.~\eqref{eq:vanishing}, Eq.~\eqref{eq:Ktot} holds at zeroth
order in $k/k_c$, with gradient terms, anisotropic stress and a
non-instantaneous angular distribution correcting it at
$\mathcal O[(k/k_c)^2]$.
In the isotropic, perfect-radiation, leading-gradient limit, this solves the
post-production evolution equation of Ref.~\cite{BS2026},
\begin{equation}
\dot K + 2HK = \frac{4}{3}H\sigma_\perp^2 ,
\label{eq:Keq}
\end{equation}
which follows from the definition \eqref{eq:Kdef} with the Raychaudhuri and
Hamiltonian constraints, independently of any matching prescription. Both
treatments evolve this same equation; the physical issue is how the production
history fixes the initial data of the total constrained system and its cross
correlation, rather than the condition assigned to $K_\perp$ in isolation.
Equation~\eqref{eq:Ktot} solves it, as is readily checked: with $\sigma_\perp^2 = S a^{-4}$ one has
$\dot K_{\rm tot} = \tfrac83 H\sigma_\perp^2$ and
$2HK_{\rm tot}=-\tfrac43 H\sigma_\perp^2$. The total $K$ thus retains only the
instantaneous shear term and decays as $a^{-4}$; the slowly decaying $a^{-2}$
relic is absent.

Equation~\eqref{eq:Ktot} does not contradict Eq.~\eqref{eq:vanishing}, but
neither is it identical to it. Through the Gauss constraint the intrinsic
curvature is ${}^{(3)}\!R = 6K_{\rm tot} + 2\sigma^2 = -2\sigma_\perp^2$, with
$\sigma^2=\sigma_\perp^2$ here: nonzero after $\eta_*$ and decaying as
$a^{-4}$. The two descriptions agree on what matters, neither leaves a slowly
decaying relic.

The white noise of Ref.~\cite{BS2026} arises from a different, and actually
acausal, choice of integration constant. Its transverse piece is fixed by
Eq.~(3) of that work, which writes the increment as an integral of
$\overline{\sigma^2}$ running from the observation redshift up to an initial
$z_{\rm i}$ and carrying an explicit $(1+z)^2$ prefactor---the $a^{-2}$ scaling
of the relic. Combined with the instantaneous-production approximation
$\Omega_{\rm GW}\propto\Theta(z_*-z)$ adopted in its Sec.~III\,A, this is
Eq.~\eqref{eq:Keq} with $K_\perp=0$ on the matching hypersurface immediately
following the idealized production event, where $K_\perp$ is continuous while
$\sigma_\perp^2(a_*)$ is already nonzero. This selects
\begin{equation}
K_\perp(a) = -\frac{2}{3}\sigma_\perp^2(a)
+ K_*\left(\frac{a_*}{a}\right)^{2},
\qquad K_*\equiv\frac{2}{3}\sigma_\perp^2(a_*),
\label{eq:Kperpsol}
\end{equation}
whose second term is the homogeneous solution of Eq.~\eqref{eq:Keq}. Its
amplitude $K_*$ is nonzero on the matching hypersurface, although the full
sectoral solution satisfies $K_\perp(a_*)=0$, because the particular and
homogeneous contributions cancel there. It decays only as $a^{-2}$ and is the
relic $K_\perp^{\rm relic}(a)\equiv K_*(a_*/a)^2$ identified with large-scale
white noise, with spectrum $P_{K_\perp}^{\rm relic}$. Comparison with Eq.~\eqref{eq:Ktot} gives
\begin{equation}
K_{\rm tot} - K_\perp
= -\frac{2}{3}\sigma_\perp^2(a_*)\left(\frac{a_*}{a}\right)^{2} ,
\label{eq:cancel}
\end{equation}
that is, the compensating contribution is exactly $-K_\perp^{\rm relic}$.

The distinction is physical rather than one of bookkeeping. On the matching
hypersurface, Eq.~\eqref{eq:Kdef} with $\delta\theta(a_*)=0$ fixes the total
initial datum, $K_{\rm tot}(a_*)=\tfrac{8\pi G}{3}\delta\rho_r(a_*)=-K_*$,
but not $K_\perp$ separately. Combined with the sectoral convention
$K_\perp(a_*)=0$, this means that the response carries $K_r(a_*)=-K_*$, not
that the parent-fluid deficit vanishes. The condition $K_\perp(a_*)=0$ is thus
admissible as a definition of the isolated transverse solution; what it leaves
open is whether its homogeneous $a^{-2}$ term survives in the total constrained
system, and in the conserved case constructed here it is cancelled in the
correlated response. The relic is a memory of the isolated-sector matching, not
of the gravitational waves themselves.

Where it resides depends on the setup: initialized at $a_*$ it is a homogeneous
integration constant, while a calculation started before production would
generate it through the longitudinal and mixed dynamics of the source epoch.
Either way it is absent from a treatment that propagates the transverse sector
alone and adds the remainder as independent positive terms.

One remark is in order : the compensating term in
Eq.~\eqref{eq:cancel} is not $\tfrac{8\pi G}{3}\delta\rho_r$ at late times.
Here $\delta\rho_r$ redshifts as $a^{-4}$, whereas the compensating term decays
as $a^{-2}$, being a difference of two solutions of Eq.~\eqref{eq:Keq} with
different initial data and hence obeying the homogeneous equation.

In the language of Eq.~\eqref{eq:crossterm} the transverse contribution and the
remainder are anticorrelated in their slowly decaying parts,
$P_{K_\perp K_r} = -P_{K_\perp}^{\rm relic}[1+\mathcal O(k^2/k_c^2)]$, so the
white relic cancels; treating them as independent positive contributions misses
this solution.

What cancels is the relic, not the instantaneous term: since $\sigma_\perp^2$
carries a white spectrum, the total remains nonzero at any finite time, but the
instantaneous contribution decays as $a^{-4}$ in power relative to the $a^{-2}$
relic and is negligible by recombination for the early production epochs
considered here. What controls the late-time residual is the gradient term. Its
power begins at $\mathcal O[(k/k_c)^4]$ in the no-dipole realization, but
generically at $\mathcal O[(k/k_c)^2]$ when the momentum dipole
$q_i\propto k_i$ allowed by Eq.~\eqref{eq:compensated} is present. Equations \eqref{eq:Ktot}--\eqref{eq:cancel} show the compensated
matching to be consistent with the post-production covariant evolution; they do
not integrate the production epoch itself.

\textit{Scope and observable consequences.---}The exact cancellation in
Eqs.~\eqref{eq:vanishing} and \eqref{eq:cancel} is a deliberately minimal
counterexample, not a universal prediction. The claim is not that a realistic
source produces no scalar signal---Eq.~\eqref{eq:vanishing} holds only at
leading gradient order---but that conservation fixes the leading piece of the
response, leaving a residual that must be computed from a source model rather
than read off from an isolated transverse contribution. A collisionless
gravitational-wave component develops anisotropic stress, and a realistic
emitter may also inject scalar momentum or anisotropic stress during
production; both require an angular distribution and an unequal-time source
correlator, neither of which is fixed by the equal-time spectrum $\Om(f)$. A
microscopic second-order solution for a specified waveform would be needed to
determine the residual beyond this compensated limit.

The scope is also restricted. Compensation requires a parent fluid from which
the wave energy is drawn, and so applies to causal production inside the
horizon; for genuinely superhorizon primordial modes there is no production
event and no parent fluid, and Eqs.~\eqref{eq:compensated} and
\eqref{eq:Ktot} do not apply. Reference~\cite{BS2026} focuses on waves
generated inside the horizon, such as those from phase transitions, so our example addresses the case of interest.

The long-wavelength scaling shows how large the difference can be. For a
localized source with compensated monopoles and $q_1=0$, the causal integral
constraints give a comoving-density response starting at $\mathcal O(k^2)$ in
amplitude \cite{HuSpergelWhite,Traschen1985}, hence $\mathcal O(k^4)$ in power;
the angular distribution that decides whether the dipole is retained is not
fixed by $\Om(f)$. In either case, if the scalar anisotropic stress is regular in the
isotropized-packet limit the post-source potential is bounded and the curvature
spectrum is the regular one of Eq.~\eqref{eq:causalscaling},
$\Delta^2_{\cR}=\mathcal O(k^3)$, rather than $\Delta^2_{\cR}\propto k^{-1}$.
We state this for $\cR$ deliberately: $K$ is a convenient local scalar for
following the covariant evolution, but it is not an observable, and the
comparison with Eq.~\eqref{eq:template} is a statement about the curvature
spectrum. The conclusion is conditional---causal sources with unsuppressed
scalar anisotropic stress need not obey it \cite{DurrerSakellariadou1997}---and
what is model independent is the need to retain the response and its cross
correlation.

For orientation, normalized at the production horizon $k_c=a_*H_*$, the leading
gradient residual carries power $(k/k_c)^4$ relative to the isolated adiabatic
template in the no-dipole realization, or $(k/k_c)^2$ with the dipole
retained. QCD-scale production at $z_*\sim10^{12}$ corresponds to
$k_c\simeq1\,\mathrm{pc^{-1}}=10^{6}\,\mathrm{Mpc^{-1}}$, the horizon scale
quoted in Ref.~\cite{BS2026}, so at $k\sim10^{-3}\,\mathrm{Mpc^{-1}}$ these
ratios are $\mathcal O(10^{-36})$ and $\mathcal O(10^{-18})$. The spread between them is the point: both display the
scale separation, but the coefficient and even the leading power depend on the
source, so neither is a new bound.

Our example is thus sufficient to reject a model-independent
identification of $P_{K_\perp}$ with $k_{\rm BH}$, but not to show that every
realistic bound must weaken: incomplete compensation, or a different
isocurvature transfer function, can yield a nonzero scalar signal and may even
tighten a constraint. Existing calculations of tensor-induced scalar
perturbations likewise depend on the fluid response and on the variable being
evolved \cite{Bari2023}.

The compensated state of Eq.~\eqref{eq:entropy} is structurally analogous to a
neutrino density isocurvature mode \cite{BMT2000}, but a quantitative bound
requires treating the waves as a separate free-streaming species with its own
Boltzmann hierarchy and anisotropic-stress transfer, for example in
{\sc class} \cite{CLASS}, normalized by a production model and with any
residual adiabatic component propagated jointly. Since isocurvature and
adiabatic modes are constrained differently \cite{PlanckInflation}, the
resulting bound need not be weaker---it is simply not determined by the
equal-time spectrum alone.

\textit{Conclusion.---}Two complementary obstructions stand in the way of
identifying a white tensor contribution to $K$ with a universal passive
adiabatic mode. For a conserved causal source, analyticity allows a regular
white curvature spectrum, $P_{\cR}=\mathcal O(k^0)$, but excludes the infrared
branch $P_{\cR}\propto k^{-4}$ assumed in Eq.~\eqref{eq:template}. And the
compensated conversion constructed here yields $k_{\rm BH}=0$ with a white
relative-entropy perturbation, its covariant form removing the slowly decaying
transverse relic through a negative cross spectrum. A likelihood based on
Eq.~\eqref{eq:template} therefore constrains a matching prescription, not a
model-independent class of causal sources.

\textit{Note added.---}After the original version of this Letter was submitted,
two second-order analyses reached related conclusions for scalar acoustic
perturbations \cite{Ireland2026,Hu2026}, by a route independent of the
construction presented here.

\end{document}